\documentclass[twocolumn,english,nobibnotes,nofootinbib]{revtex4-1}
\usepackage[T1]{fontenc}
\usepackage[latin9]{inputenc}
\usepackage{geometry}
\usepackage{amsbsy}
\usepackage{graphicx}
\usepackage{babel}
\usepackage{color}
\usepackage{amsmath}
\begin{document}

\author{E. N. Osika}
\affiliation{AGH University of Science and Technology, Faculty of Physics and
Applied Computer Science,\\
 al. Mickiewicza 30, 30-059 Krak\'ow, Poland}

\author{B. Szafran}
\affiliation{AGH University of Science and Technology, Faculty of Physics and
Applied Computer Science,\\
 al. Mickiewicza 30, 30-059 Krak\'ow, Poland}

\title{Spin-valley resolved photon-assisted tunneling in carbon \\ nanotube double quantum dots}
\begin{abstract}
We consider the photon-assisted tunneling (PAT) and the Landau-Zener-Stueckelberg (LZS) interference for double quantum dots induced electrostatically along
a semiconducting carbon nanotube. An atomistic tight-binding approach and the time-dependent configuration interaction method are employed
for description of the systems of a few confined electrons and holes. 
We reproduce the patterns of the LZS interference recently observed for the quantum double dots describing transport across hole-localized states.
Moreover, we indicate that for charge configurations for which the ground-state is Pauli blocked PAT can be used for resolution of the 
transitions that involve spin-flip or intervalley transitions without the spin-valley conserving background signal.
\end{abstract}
\maketitle
\section{Introduction}

Photon assisted tunneling   (PAT) 
across quantum dots (QD) \cite{pat,stafford,blick} is observed in microwave fields  when the Fermi level electrons pass across the higher-energy confined levels 
upon absorption of the energy from the radiation field. 
The phenomenon  has a resonant character and occurs provided that the microwave frequency is in resonance with the mismatch between the energy levels $\Delta E$ so that the electron absorbs
a single or $n$ photons in order to climb the higher energy level, $n\hbar\omega =\Delta E$.
PAT has beed used for spectroscopy of the dot-confined energy levels \cite{oster, meyer,blick}
and charge dynamics in multiple quantum dots \cite{bao,shang,petersson}.
PAT in double quantum dots has also beed used for the spin-related phenomena 
including transport involving spin flips  \cite{schreiber} and  manipulation of the spin qubits
 \cite{petta,ko}. For strong microwave fields within double quantum dots the PAT enters the regime of the Landau-Zener-Stueckelberg (LZS) \cite{shevchenko,stehlik} interference
when the system is driven by the ac electric field across the avoided crossing between energy levels of different charge occupation. The procedures for fast spin-flips based on the LZS interference were proposed \cite{pes,gau,shang}.
The LZS interference pattern has also been used for studies of the dephasing processes \cite{mavalankar,shevchenko} in double dots, including the spin coherence times \cite{oliver,sili},
as well as for sensitive residual radiation detectors \cite{dovinos}, charge \cite{prati} and spin pumping \cite{bu}.

The PAT and LZS interference were also observed in quantum dots defined in semiconducting carbon nanotubes (CNT) \cite{meyer,mavalankar}. 
In CNTs - which are graphene related material with strong spin-orbit coupling due to the curvature of the carbon plane \cite{laird,ando,so1,so2,so3,so4}
--  in addition to the spin degree of freedom the valley one is present. In this paper we consider
detection of the  photon assisted tunneling involving spin flips and intervalley transitions. The study is based
on the time dependent configuration interaction approach for systems of several carriers described within the atomistic tight-binding 
approach. The study covers simulation of the LZS
interference pattern as recently observed \cite{mavalankar}. Moreover, we indicate that for systems in which the charge transport
is blocked by the Pauli blockade the LZS pattern contains clear separate lines corresponding to either the spin or the valley flips
accompanying the electron hopping.


\section{Theory}

\begin{figure}[htbp]
 \includegraphics[width=0.6\linewidth]{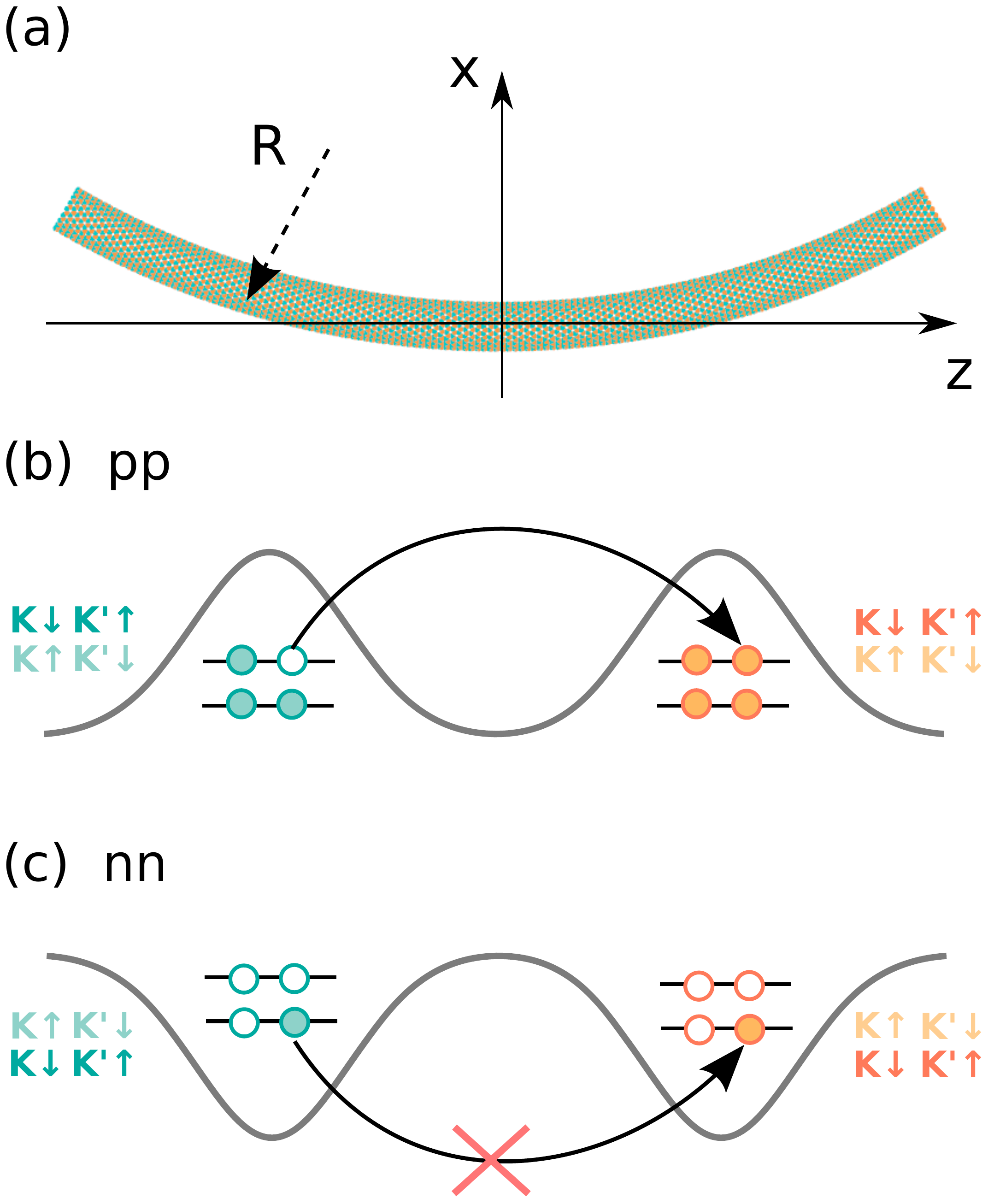} 
\caption{(a) Scheme of the system considered in the paper. (b) (1h,0) and (c) (1e,1e) charge configuration of the $pp$ and $nn$ quantum dots, respectively. The arrows indicate (b) the (1h,0)$\rightarrow$(0,1h) and (c) the (1e,1e)$\rightarrow$(0,2e) transition. 
} \label{schemat}
\end{figure}

We model a semiconducting CNT of length $L=53.11$ nm, diameter $2r=1.33$ nm and chiral vector $C_h=(17,0)$. We consider both a straight nanotube
 and a bent \cite{flensb} one with radius of the bent $R=40$ nm [see Fig.\ref{schemat}(a)]. The nanotube is suspended above the electrostatic gates which produce a double quantum dot confining potential
\begin{equation}
 W_{QD}(z)= V_1 e^{-(z+z_s)^2/d^2}+V_2 e^{-(z-z_s)^2/d^2}.
\end{equation}
In Eq. (1) $d$ determines the QD widths, $z_s$ is a shift of the dots from $z=0$ and $V_1$/$V_2$ is a potential of the left/right QD. We consider nanotube in $pp$ and $nn$ configurations in which both the dots are occupied either by holes or excess electrons, respectively. For a $pp$ configuraton we apply $V_1=V_p$, $V_2=V_p-\Delta$, and for the $nn$ configuration $V_1=V_n-\Delta$, $V_2=V_n+\Delta$, where $\Delta$ defines energy mismatch between potentials on the left and right dot. In the calculations we apply $V_p=0.3$ eV, $V_n=-0.5$ eV, $z_s=10$ nm and $d=4$ nm. We apply an external magnetic field of magnitude $B$ parallel to the $z$ direction.

We calculate single-electron states using the tight-binding approximation with the $p_z$ orbitals. Hamiltonian of the eigenproblem to be solved, reads
\begin{eqnarray}\label{ham1e}
&H=&\sum\limits_{\{i,j,\sigma,\sigma'\}}(c_{i\sigma}^{\dagger}t_{ij}^{\sigma\sigma'}c_{j\sigma'}+h.c.)+ \\ &\sum\limits_{i,\sigma,\sigma'}c_{i\sigma}^{\dagger}&\left(W_{QD}({\bf r}_{i})\delta_{\sigma\sigma'}+\frac{g_L\mu_{b}}{2}\boldsymbol{\sigma}^{\sigma\sigma'}\cdot{\bf B}\right)c_{i\sigma'}\nonumber.\label{ham}
\end{eqnarray}
The first sum in Eq. (\ref{ham1e}) accounts for the hopping between the nearest neighbor atoms and the second sum for the interaction with the external electric and magnetic fields. Here $c_{i\sigma}^{\dagger}\ensuremath{}(c_{i\sigma})$ is the particle creation (annihilation) operator at ion $i$ with
spin $\sigma$, $t_{ij}^{\sigma\sigma'}$ is the spin-dependent hopping parameter,  $\delta_{\sigma\sigma'}$ stands for the Kronecker delta, $g_L=2$ for the Land\'e factor, $\mu_{b}$ the Bohr magneton and $\boldsymbol{\sigma}$ for the vector of Pauli matrices. The hopping parameters $t_{ij}^{\sigma\sigma'}$ contain the spin-orbit interaction which arises from the curvature of the graphene plane \cite{laird,ando,so1,so2,so3,so4}. 
Both the folding of the graphene plane into the tube and the bent of the nanotube \cite{JPCM} are accounted for. Moreover, $t_{ij}^{\sigma\sigma'}$ include Peierls phase which accouts for the orbital magnetic moments interaction with the external magnetic field. The explicit form of the hopping parameters has been given in Ref. \cite{JPCM}.

We calculate few electron eigenstates using the configuration-interaction (CI) method. We solve the eigenproblem of the Hamiltonian 
\begin{equation}
H_{CI}=\sum_{a}\epsilon_{a}g_{a}^{\dagger}g_{a}+\frac{1}{2}\sum_{abcd}V_{ab;cd}g_{a}^{\dagger}g_{b}^{\dagger}g_{c}g_{d}, \label{CI}
\end{equation}
where $\epsilon_{a}$ is the energy of the $a$-th eigenstate of Hamiltonian $H$, $g_{a}^{\dagger}$ and $g_{a}$ are the creation and annihilation operators of the electron in the $a$-th state and $V_{ab;cd}$ are electron-electron interaction  matrix elements (with the form given in Ref. \cite{PRBambi}).\\

For $pp$ quantum dots we focus on a (1h,0) charge configuration (single hole localized in the left dot) and its transition to (0,1h) state (single hole localized in the right dot). To model the system we consider seven last electrons of the two highest valence band orbitals [see Fig.\ref{schemat}(b)] confined in the left and right dot, each of the orbitals nearly fourfold degenerate with respect to the valley and the spin -- i.e. the system with a single unoccupied state of the valence band.
For $nn$ quantum dots we study  the charge configurations for which Pauli blockade appears, i.e. (1e,1e) (one electron in each dot) and its transition to (0,2e) (two electrons in the right dot). For that transition we consider 2 electrons at the bottom of the conduction band [see Fig.\ref{schemat}(c)] and include the 8 lowest conduction band orbitals (32 states) in the CI basis.  

The dynamics of the system driven by external ac field is simulated by solving the time-dependent Schr\"{o}dinger equation. The time-dependent Hamiltonian reads
\begin{equation}
H'(t)=H_{CI} + \sum_{j=1}^N eF z_j \sin(\omega t),
\end{equation}
where $F$ is the ac electric field amplitude, $\omega$ is its frequency, $N$ is a number of electrons (7 for $pp$ and 2 for $nn$ quantum dots). We solve the time-dependent Schr\"{o}dinger equation within the basis of the eigenstates $\Psi_n$ of  Hamiltonian $H_{CI}$
\begin{equation}
\Psi({\bf r}_{1...N}, {\boldsymbol \sigma_{1...N}}, t) =  \sum_n c_n(t) \Psi_n({\bf r}_{1...N}, {\boldsymbol \sigma_{1...N}}) e^{-\frac{iE_nt}{\hbar}}.
\end{equation}
In this basis the Schr\"{o}dinger equation $i\hbar\frac{\partial\Psi}{\partial t}=H'\Psi$ takes the form
\begin{equation}
i\hbar \dot{c}_k(t)=\sum_n c_n(t) eF \sin(\omega t) \langle\Psi_k|z|\Psi_n\rangle e^{-\frac{i(E_n-E_k)t}{\hbar}}. \label{cndot}
\end{equation}
We discretize equation (\ref{cndot}) in time using the Crank-Nicolson algorithm.

\section{Results}
\subsection{Photon assisted (1h,0)$\rightarrow$(0,1h) transitions}

\begin{figure}[htbp]
 \includegraphics[width=0.95\linewidth]{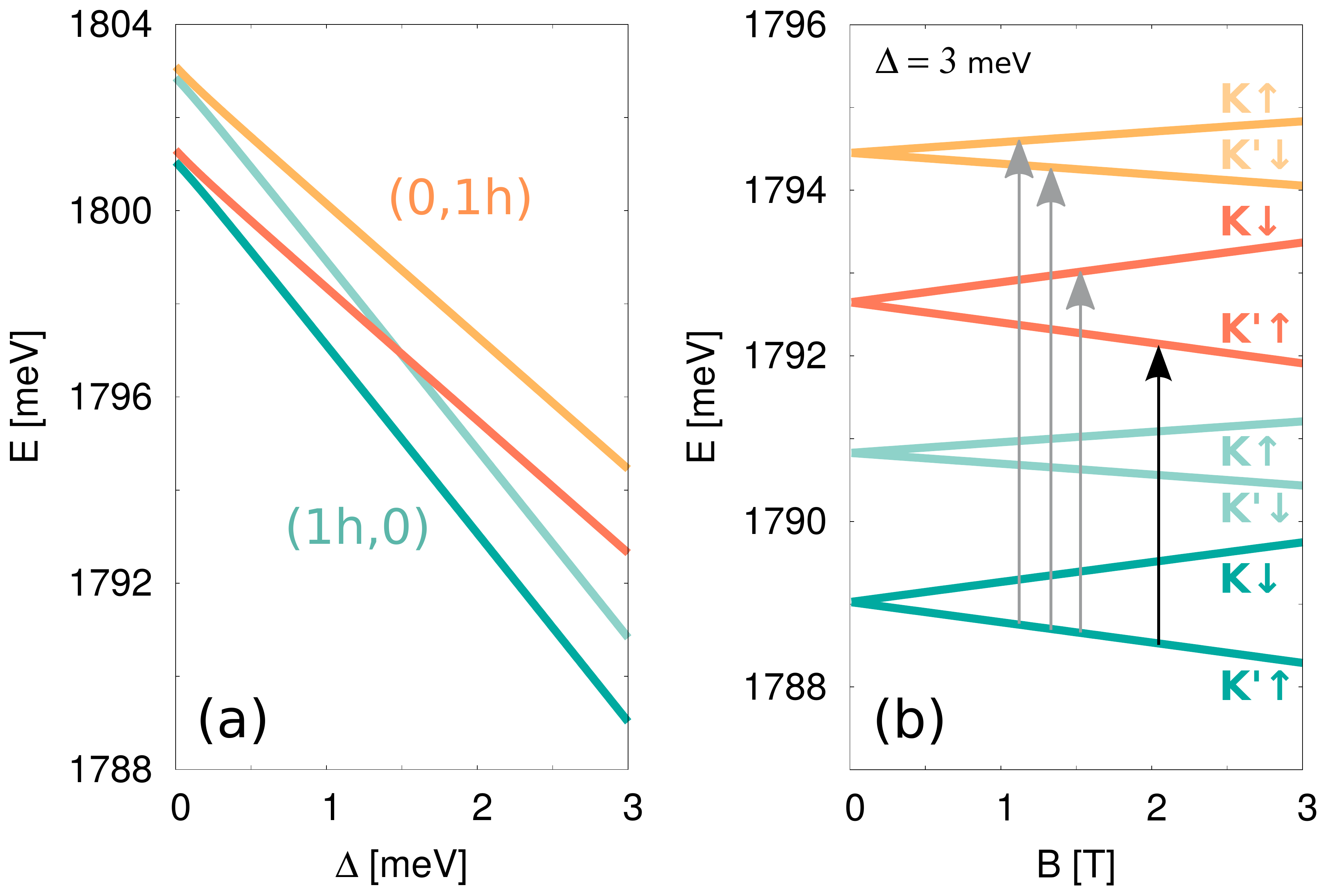} 
\caption{(a) Single-hole energy levels in a $pp$ QD as a function of the potential difference $\Delta$ between the dots. (b) Energy levels dependence on the external magnetic field $B$ for $\Delta=3$ meV. The symbols $K/K'$~$\uparrow/\downarrow$ indicate the spin-valley states of the hole [the single unoccupied energy level of the octet at the top of the valence band -- cf. Fig. 1(b)] and the arrows represent possible (1h,0)$\rightarrow$(0,1h) transitions.
} \label{widmo_deltaB}
\end{figure}

\begin{figure}[htbp]
 \includegraphics[width=0.95\linewidth]{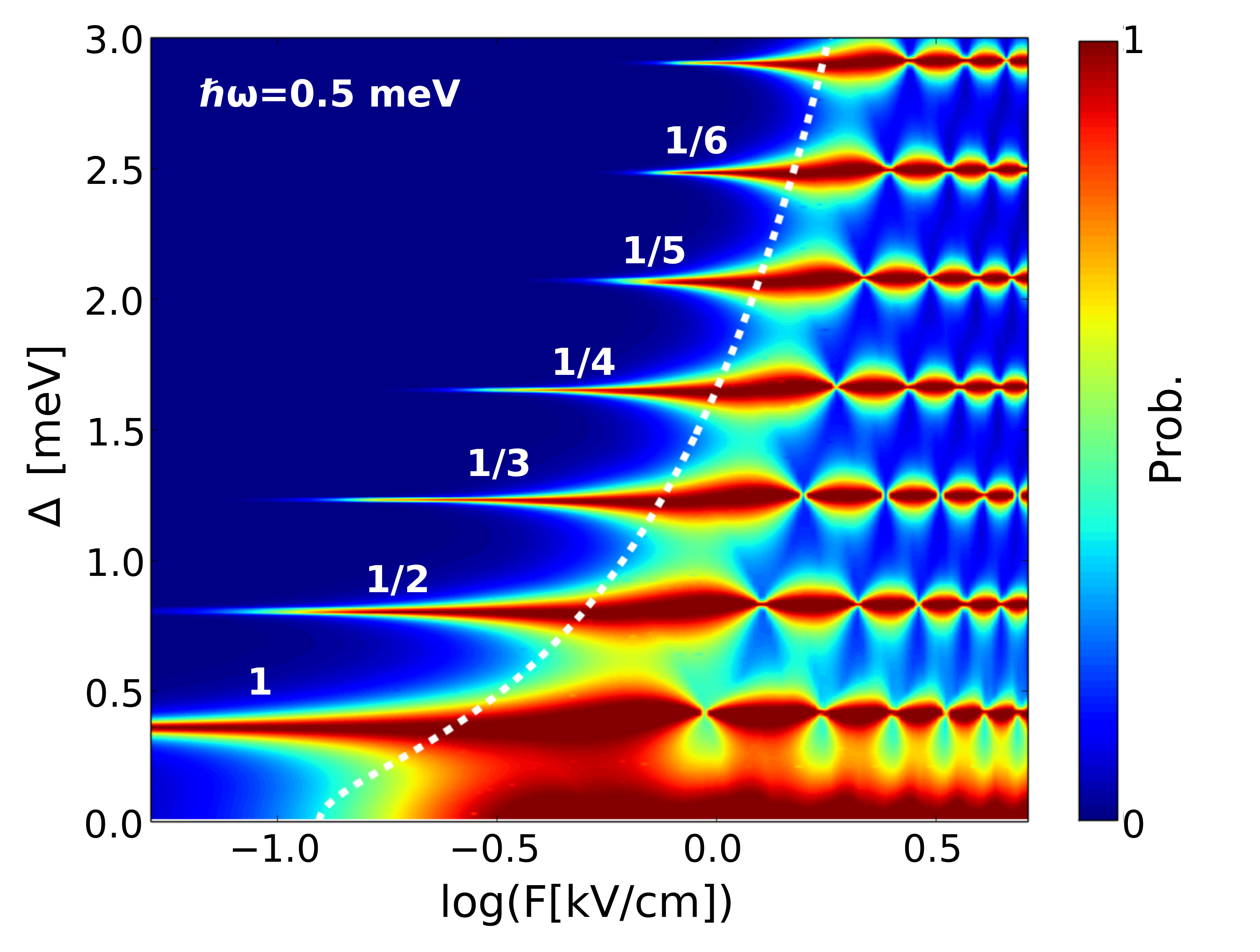} 
\caption{Maximal  probability of  (1h,0)$K'\uparrow$~$\rightarrow$~(0,1h)$K'\uparrow$ transition for straight and clean CNT as a function of $\Delta$ and a logarithm of the amplitude of the microwave electric field $F$. We apply the frequency $\hbar\omega=0.5$~meV within a pulse of duration 1 ns. White numbers on the plot indicate the multiphoton resonances at the 1, 1/2, 1/3 etc. of the nominal resonant frequency for the (1h,0)$K'\uparrow$~$\rightarrow$~(0,1h)$K'\uparrow$ transition. The thick dashed line shows the electric field required to take the system to the center of the avoided
crossing between the (1h,0) and (0,1h) energy levels for a given detuning $\Delta$.
} \label{skan_prosta}
\end{figure}

\begin{figure}[htbp]
 \includegraphics[width=0.85\linewidth]{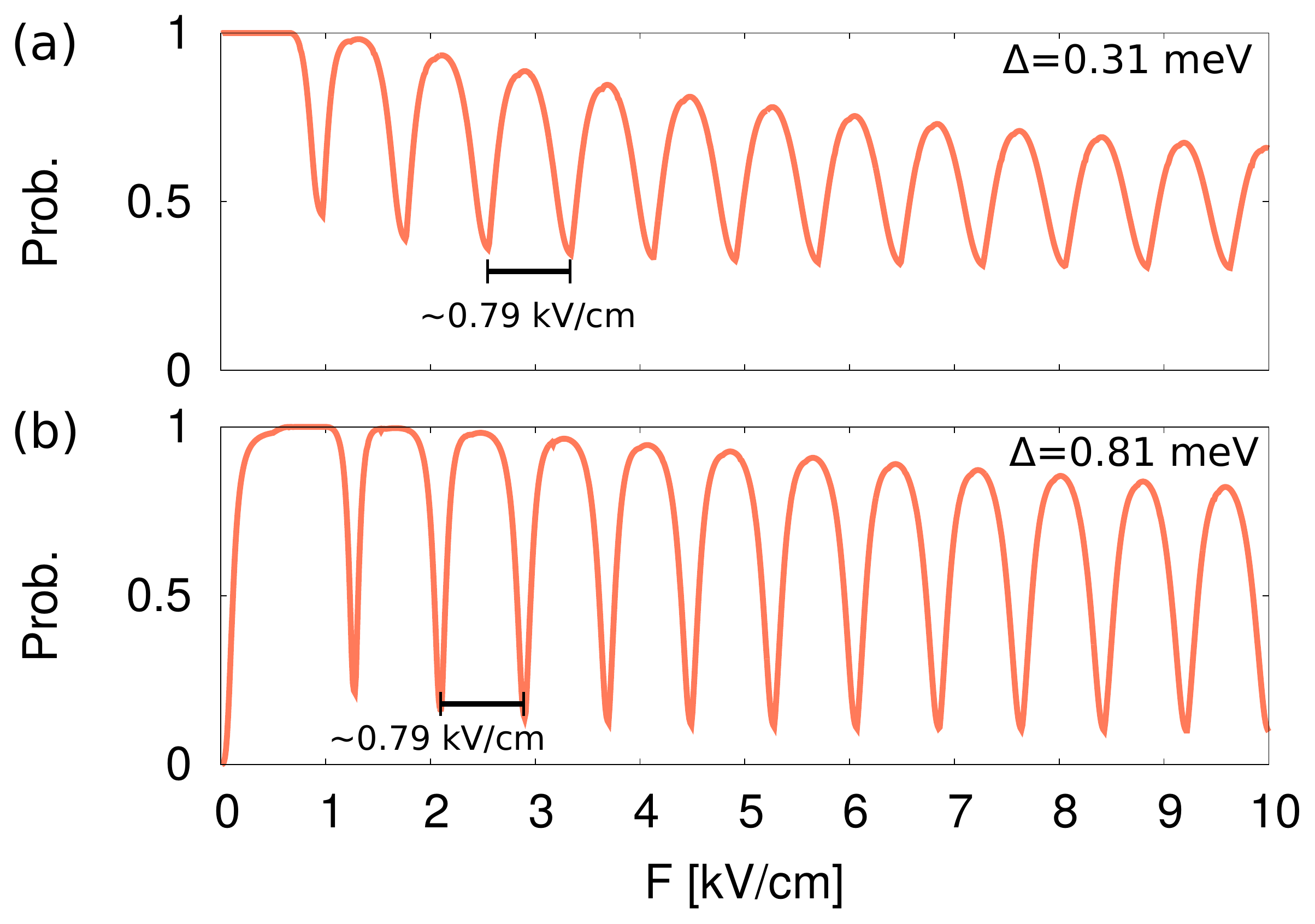} 
\caption{Cross-sections of Fig.~\ref{skan_prosta} for (a) $\Delta=0.31$ meV [direct (1h,0)$K'\uparrow$~$\rightarrow$~(0,1h)$K'\uparrow$ transition] and (b) $\Delta=0.81$ meV (two-photon transition). } \label{LZS}
\end{figure}

In Fig. \ref{widmo_deltaB}(a) we present the single-hole lowest-energy levels of $pp$ system as a function of the potential difference $\Delta$ between the dots. For equal potentials on both dots ($\Delta=0$) the hole in the lowest energy states occupies both of the dots evenly. Note, that the levels of the (1h,0) and (0,1h) branch are non-degenerate at $\Delta=0$. The avoided crossing of width 0.249 meV  due to the tunnel coupling between the dots will be used in simulation of the LZS interference.
For positive $\Delta$ the (1h,0) charge configuration is promoted to the ground state. Each line in Fig. \ref{widmo_deltaB}(a) is twofold degenerate and the energy splitting of about 2 meV between the pairs of blue/orange lines originates from the SO interaction. The degeneracy is lifted by the external magnetic field $B$, as shown in Fig. \ref{widmo_deltaB}(b).  The symbols $K/K'$, $\uparrow/\downarrow$ indicate the valley and spin states of the hole (the only empty state in the valence band). 

We focus on the transitions between (1h,0) states (blue lines in Fig. \ref{widmo_deltaB}) and (0,1h) states (orange lines). These can be understood as the hole jumping from the left dot to the right one. We assume that  the hole initially occupies the $K'\uparrow$ ground state. In straight, clean carbon nanotube the hole can tunnel to the right dot only by the transition to the same spin and valley state [black arrow in Fig. \ref{widmo_deltaB}(b)]. Transition to the states of different spin and/or valley [grey arrows in Fig. \ref{widmo_deltaB}(b)]
 are possible provided the symmetry of the nanotube is broken: the spin-orbit coupling itself does not allow for the spin transitions \cite{eosi1} and a short range defect is needed to drive the 
valley flips \cite{pal}. 
Mixing of the spin or valley degree of freedom can be obtained by bending the nanotube or introducing disorder in the lattice, respectively.

In Fig. \ref{skan_prosta} we plot the probability of the transition (1h,0)$\rightarrow$(0,1h) in a straight and clean CNT as a function of $\Delta$ and a logarithm of the amplitude of the microwave field $F$. The microwave frequency of $\hbar\omega=0.5$~meV was assumed and the simulation time covered 1 ns. In Fig. 3 several transition lines can be observed which correspond to the direct transition from (1h,0)~$K'\uparrow$ state to (0,1h)~$K'\uparrow$ state (noted as ``1'' in Fig. \ref{skan_prosta}) and its harmonics (``1/2``, ''1/3``, etc.) for multiphoton transitions \cite{stehlik}.
The dashed line indicates the axial electric field that is necessary to reach the center of the avoided crossing of the (1h,0), (0,1h) configurations  for a given $\Delta$, i.e. $F= \frac{E_{(0,1h)}(\Delta)-E_{(1h,0)}(\Delta)}{2ez_s}$.
To the right of the dashed line the electric field drives the system across the (1h,0), (0,1h) avoided crossing which allows for observation of the LZS interference. 
A clear interference pattern appears to the right of this line in Fig.~\ref{skan_prosta}.  Together with the periodic variation of the electric filed $eF z \sin(\omega t)$ the system goes back and forth across the (1h,0)$\leftrightarrow$(0,1h) avoided crossing. Each passage through the avoided crossing region results both in non-adiabatic transition between the states and in acquisition of the additional phases by them - different for (1h,0) and (0,1h) states. Depending on the relative phase acquired by the system during the time evolution we observe either constructive or destructive interference. It has been shown in Ref.~\cite{shevchenko} that within slow-passage limit the position of the minima/maxima of the probability is proportional to $F/\omega$ ratio. In Fig.~\ref{LZS} we plot the cross-sections of Fig.~\ref{skan_prosta} for (a) $\Delta=0.31$ meV and (b) $\Delta=0.81$ meV with linear $F$ scale. 
Both for direct [Fig.~\ref{LZS}(a)] and two-photon [Fig.~\ref{LZS}(b)] transition we indeed observe equal distances of about 0.79 kV/cm on $F$ scale between the consecutive extrema of the probability, which confirms the LZS origin of the observed oscillations. Since away from $\Delta=0$ the energy levels depend linearily on $\Delta$ [see Fig. \ref{widmo_deltaB}(a)], the distance between the adjacent harmonics on the $\Delta$ scale is constant. The results of Fig. 3 agree with the ones obtained in the experiment   of Ref. \cite{mavalankar}.

\begin{figure*}[htbp]
 \includegraphics[width=0.85\linewidth]{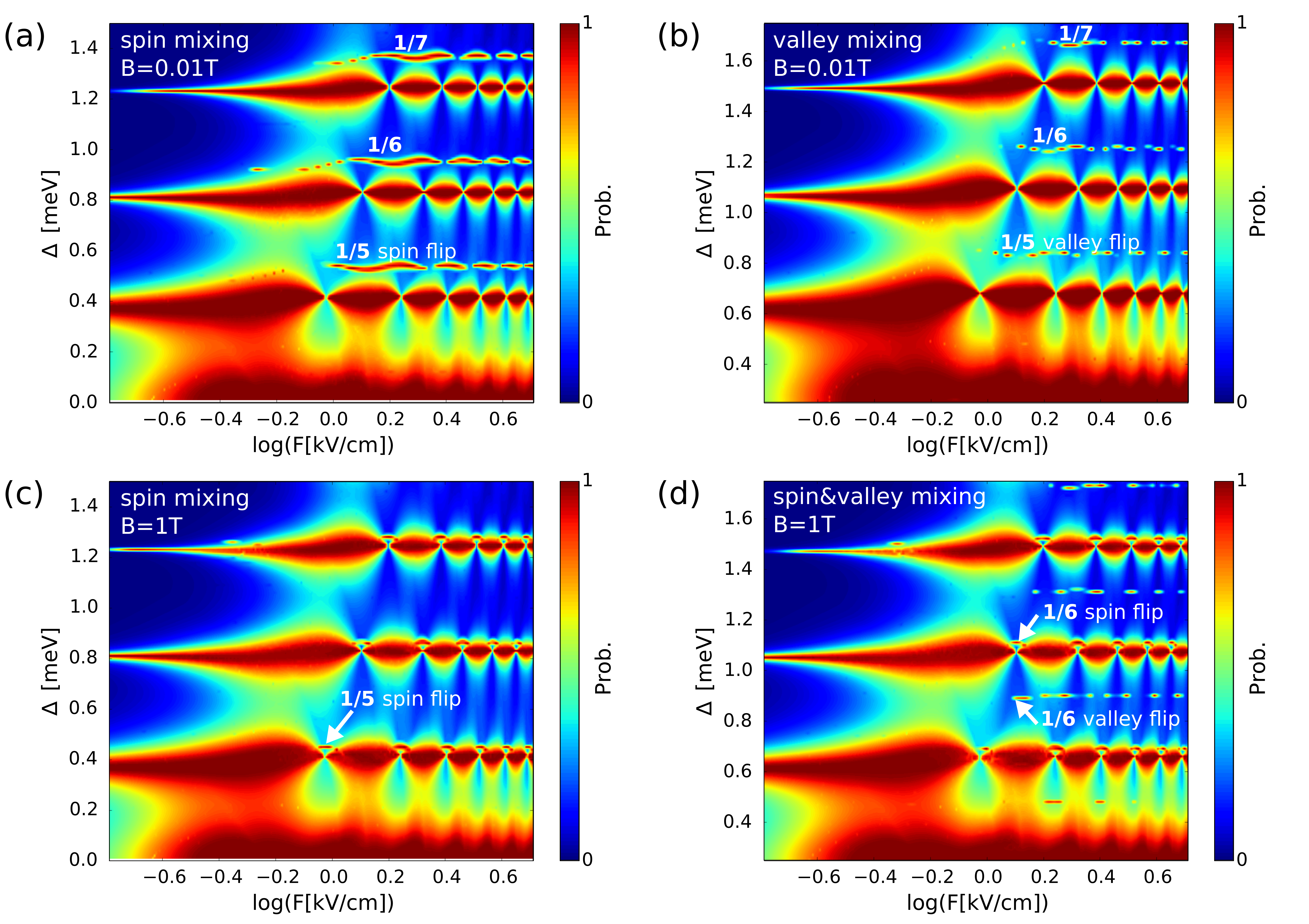} 
\caption{Maximal probability of the (1h,0)~$\rightarrow$~(0,1h) transitions  as a function of $\Delta$ and a logarithm of the amplitude of the microwave electric field $F$. We apply ac electric field frequency $\hbar\omega=0.5$~meV within a pulse of duration 1 ns. In (a) and (c) the nanotube is assumed bent, in (b) the two potential peaks of 1 eV at -14.2 nm and 14.13 nm are introduced in the CNT, in (d) both the bending and the potential peaks are considered. In (a,b) the magnetic field is set close to zero ($B=0.01$ T), in (c,d) $B=1$ T. } \label{skany}
\end{figure*}

In Fig.~\ref{skany}(a) we reproduce the results of Fig.~\ref{skan_prosta} for a bent CNT. The bending mixes the spins and allows for the hole transition from one dot to the other accompanied by the spin flip. The lines appearing due to the bending correspond to the transition from (1h,0)~$K'\uparrow$ state to (0,1h)~$K'\downarrow$ state at 1/5-th, 1/6-th and 1/7-th of the nominal resonant frequency. The distance between the harmonics on the $\Delta$ scale is constant since the energy difference between the two states participating in the transition is linear with $\Delta$ [see Fig.~\ref{widmo_deltaB}(a)]. Moreover, the $\Delta$ distances between adjacent harmonics are the same for spin-conserving and spin-flipping transitions. However, the lines corresponding to these two transitions are shifted with respect to each other due to the SO splitting of the spin-valley doublets of  2 meV which is constant for every $\Delta$ -- see Fig. 2(a).

To enable the valley-flipping transitions -- i.e. (1h,0)~$K'\uparrow$ $\rightarrow$ (0,1h)~$K\uparrow$ -- a short-range potential has to be introduced in the CNT lattice. That type of potential acts as a scattering centre which mixes $K$ and $K'$ valleys. In experimental setups the scattering potential might by produced by defects in the lattice. To model the effect in our calculations we introduce 2 potential peaks of 1~eV at -14.2 nm and 14.13 nm. That results in additional narrow resonances visible in Fig.~\ref{skany}(b) as opposed to Fig.~\ref{skan_prosta}. Similarly to Fig.~\ref{skany}(a), the lines correspond to the 1/5-th, 1/6-th and 1/7-th of the nominal resonant frequency for the transition (1h,0)~$K'\uparrow$ $\rightarrow$ (0,1h)~$K\uparrow$. The valley-flipping transitions are shifted with respect to the valley-conservant ones the same way as transitions of Fig.~\ref{skany}(a). In fact, if both spin and valley mixing are introduced, the resonant lines for spin- and valley-flipping transitions overlap. This can be easily understood looking at Fig. \ref{widmo_deltaB} -- for $B=0$ the (0,1h)~$K'\downarrow$ and $K\uparrow$ are degenerate.

The position of the resonant lines for spin/valley-flipping and spin-valley-conserving transitions with respect to each other depends on the SO energy splitting and the applied ac field frequency. While the SO interaction due to the curvature of the carbon plane is hardly controllable from an experimental point of view, the microwave driving frequency can  easily be modulated. For instance by doubling the $\omega$ we will obtain two times larger gaps between adjacent harmonics on the $\Delta$ scale. Another externally tunable parameter which can be used to modify the position of transition lines is the magnetic field $B$. Distinct behavior of the different spin and valley states in external magnetic field [see Fig.~\ref{widmo_deltaB}(b)] results in changes of the resonant frequencies for spin- and valley-flipping transitions. The spin-valley-conserving transition preserves the same resonant frequency for every $B$ since the energy difference between same spin and valley states of (1h,0) and (0,1h) configurations does not change in magnetic field. In Fig.~\ref{skany}(c) we present spectra for the bent CNT with magnetic field of $B=1$~T applied. Here, the narrow spin-flipping resonances of Fig.~\ref{skany}(a)  have shifted to lower $\Delta$ while spin-valley-conserving transition lines have remained still. As it is shown in Fig.~\ref{skany}(c), the spin-flipping resonances can be shifted until they overlap with the wider spin-valley-conserving ones and the two become nearly indistinguishable. The narrow resonances can be observed as a small perturbation in the -- otherwise regular -- interference pattern of the wide ones.

In Fig.~\ref{skany}(d) we plot the transition spectra for a bent nanotube with the potential peaks introduced in the lattice and magnetic field of $B=1$ T applied. Here still the spin-valley-conserving and spin-flipping resonances overlap but additional lines of valley-flipping transitions appear, as in Fig.~\ref{skany}(b). Note, however, that these lines are strongly shifted by the magnetic field -- for $\Delta$ of about 0.9 meV we observe 5-th harmonic for $B\approx 0$ and 6-th harmonic for $B=1$~T. The valley-flipping transition lines are shifted more than the spin-flipping ones because the energy difference between (1h,0)~$K'\uparrow$ and (0,1h)~$K\uparrow$ grows faster with $B$ then between (1h,0)~$K'\uparrow$ and (0,1h)~$K'\downarrow$ [see Fig.~\ref{widmo_deltaB}(b)].

\subsection{(1e,1e)$\rightarrow$(0,2e) transitions}

\begin{figure}[htbp]
 \includegraphics[width=0.95\linewidth]{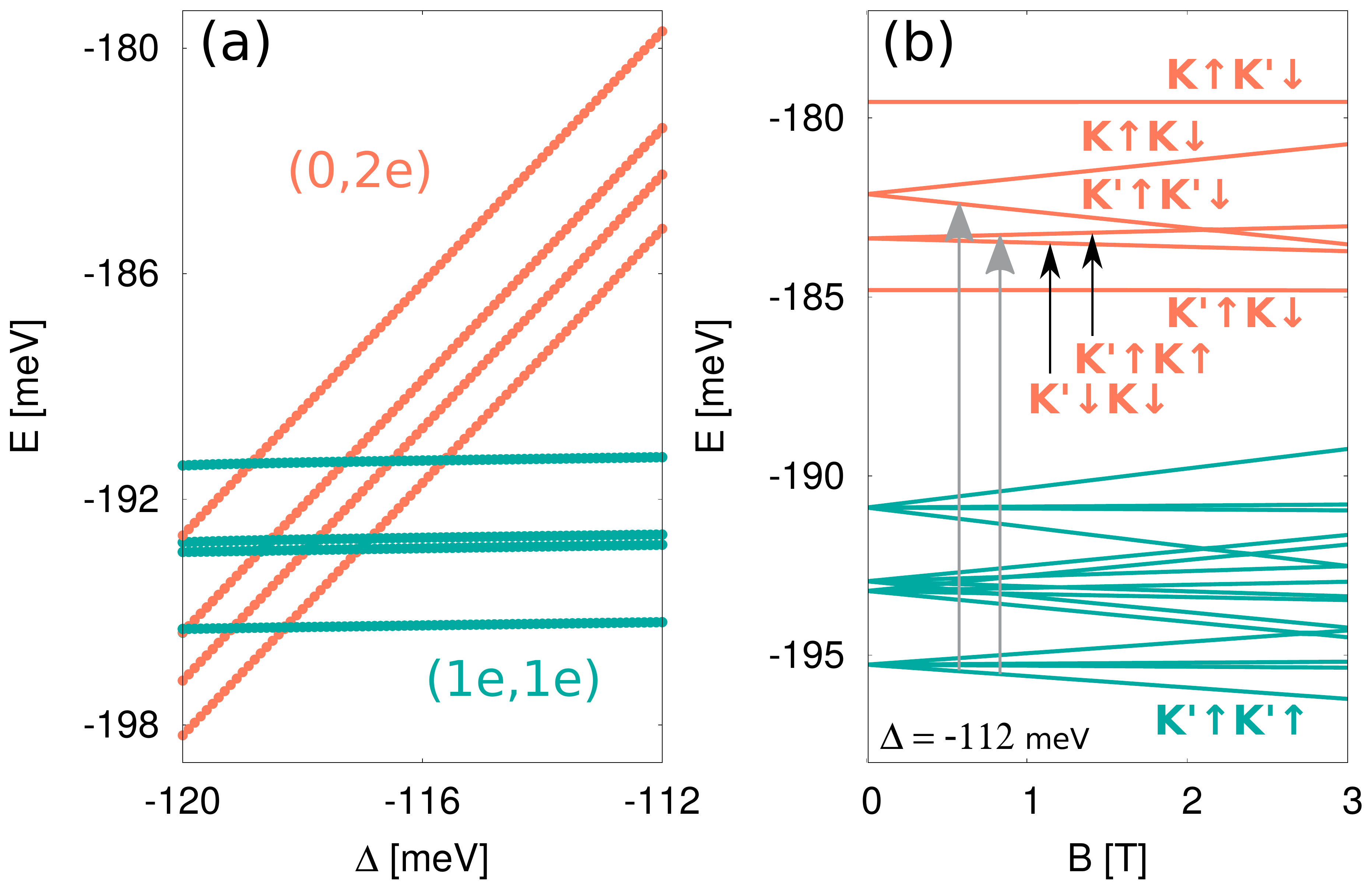} 
\caption{(a) Two-electron energy levels in $nn$ QD as a function of $\Delta$. (b) Energy levels dependence on the external magnetic field $B$ for $\Delta=-112$ meV. } \label{widmo_nn}
\end{figure}

The two-electron $nn$ system provides entirely different frame for the photon-assisted tunneling. The reason is the Pauli blockade which arises in the ground state of (1e,1e) charge configuration in the $nn$ quantum dot in the external magnetic field  [see Fig.~\ref{schemat}(c)]. In Fig.~\ref{widmo_nn}(a) we present the lowest-energy levels of two-electron $nn$ system as a function of $\Delta$ and in Fig.~\ref{widmo_nn}(b) the effect of the magnetic field $B$ on the energy levels. We are interested in transitions (1e,1e)$\rightarrow$(0,2e), i.e. tunneling of an electron from  the left dot to the right one. 
As one can  see in Fig.~\ref{widmo_nn}(b), in nonzero magnetic field the two electrons in the quantum dots occupy the same spin and valley states (triplet state $K'\uparrow K'\uparrow$) hence the tunneling from one dot to the other is suppressed. In fact, for straight and clean CNT we do not observe any transitions (1e,1e)$K'\uparrow K'\uparrow$~$\rightarrow$~(0,2e) driven by a microwave electric field. Resonant lines analogous to Fig.~\ref{skan_prosta} do not appear in the spectra if the spin-valley blockade persists in the microwave radiaton.

\begin{figure}[htbp]
 \includegraphics[width=0.95\linewidth]{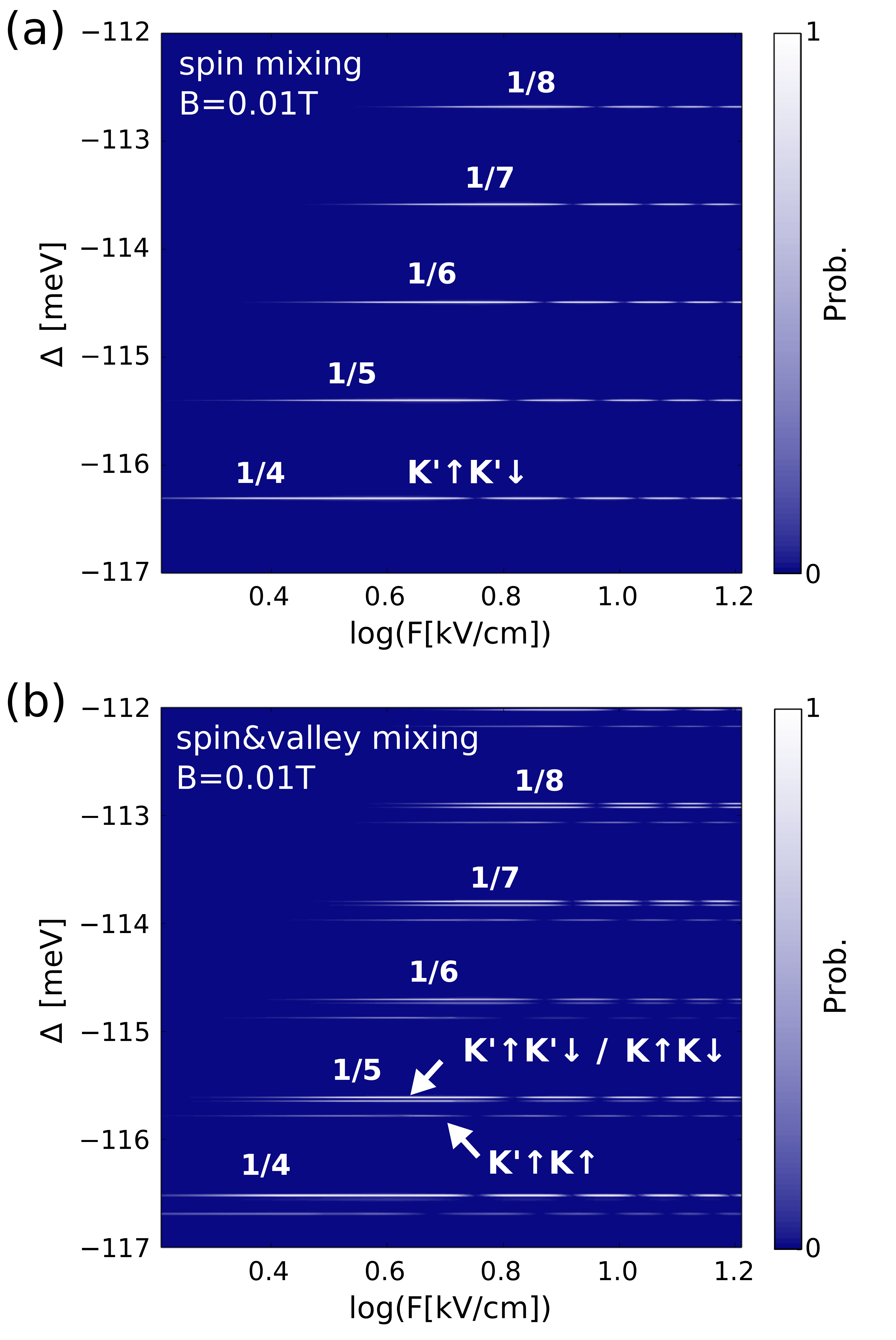} 
\caption{Maximal probability of the transition (1e,1e)~$K'\uparrow K'\uparrow \rightarrow$~(0,2e) for a bent CNT as a function of $\Delta$ and a logarithm of the amplitude of the microwave electric field $F$. 
In (a)  the nanotube is assumed bent, in (b) the bend and the two potential peaks of 1 eV at -14.2 nm and 14.13 nm are considered. We apply the electric field frequency $\hbar\omega=1.5$~meV within a pulse of duration 1 ns. The final state of the (0,2e) charge configuration is marked in the figure.
} \label{skan_nn}
\end{figure}

However, the blockade can be lifted by mixing the spin or/and valley states which can be achieved -- similarly to the $pp$ system -- by bending the nanotube or introducing defects in the lattice. In Fig.~\ref{skan_nn}(a) we plot the (1e,1e)~$\rightarrow$~(0,2e) transition probability for a bent CNT as a function  of $\Delta$ and a logarithm of the amplitude of the microwave electric field $F$. 
We assume the system is initially in the blocked triplet  $K'\uparrow K'\uparrow$ state. The only resonances that appear in Fig.~\ref{skan_nn}(a) are very narrow transition lines 
to the singlet state (0,2e)~$K'\uparrow K'\downarrow$.

Figure \ref{skan_nn}(b) contains the result in presence of the valley mixing short-range potentials that allow for the valley flips along the tunneling. Now, the lines split in double which correspond to either
spin flip (1e,1e)$K'\uparrow K'\uparrow \rightarrow$(0,2e)$K'\uparrow K'\downarrow$ or the valley flip (1e,1e)$K'\uparrow K'\uparrow \rightarrow$(0,2e)$K'\uparrow K\uparrow$. Additionaly, the line (1e,1e)$K'\uparrow K'\uparrow \rightarrow$(0,2e)$K'\uparrow K'\downarrow$ also splits into two, which results from the mixing of the $K'\uparrow K'\downarrow$ and $K\uparrow K\downarrow$ states which is present  for weak $B$. Therefore, two separate spin transition lines are visible in Fig.~\ref{skan_nn}(b) very close on the $\Delta$ scale. Other transitions, involving spin and valley transition or the change of both occupied single-electron orbitals are too weak to be noticed in this plot.

\section{Conclusions}

We have simulated the photon-assisted tunneling for a double quantum dot system defined within a carbon nanotube using the tight-binding approach and the time dependent configuration
interaction method. We considered unipolar quantum dots confining either holes or electrons for systems in which the photon induced charge transition from the ground-state to the excited state 
is either allowed or blocked by the Pauli exclusion principle. For the former case we reproduced the pattern of the LZS interference recently observed  \cite{mavalankar}. In the latter case 
in external magnetic field the ground-state of the (1e,1e) charge configuration is a nondegenerate  triplet-like spin-valley polarized state [Fig. 6(b)] $K'\uparrow K'\uparrow$
from which the charge transition to (0e,2e) can only occur provided that either the spin or the valley change at the tunneling. 
The system can be tuned by voltages into a regime where only the  ground-state is below the Fermi energy of the drain. Then, the photon assisted tunneling can be used to trigger 
and resolve charge hoppings involving either the spin or the valley transitions. 

\section*{Acknowledgments}
This work was supported by the National Science Centre
according to decision DEC-2013/11/B/ST3/03837. 
Calculations were performed in the PL-Grid Infrastructure.

\end{document}